\documentclass[final,times,twocolumn,sort&compress]{elsarticle}
\usepackage{amsmath}
\usepackage{amssymb}
\usepackage{graphics}

\journal{Physics Letters A}

\begin{document}
 
\begin{frontmatter}

\title{Absence of a spontaneous long-range order in a mixed spin-(1/2, 3/2) Ising model on a decorated square lattice due to anomalous spin frustration driven by a magnetoelastic coupling\tnoteref{grant}} 
\author[upjs]{Jozef Stre\v{c}ka}
\ead{jozef.strecka@upjs.sk}
\author[ufla]{Onofre Rojas} 
\author[ufla]{Sergio Martins de Souza}
\address[upjs]{Department of Theoretical Physics and Astrophysics, Faculty of Science, P. J. \v{S}af\'{a}rik University, Park Angelinum 9, 040 01 Ko\v{s}ice, Slovakia}
\address[ufla]{Departamento de F\'isica, Universidade Federal de Lavras, CP 3037, 37200-000, Lavras-MG, Brazil}
\cortext[grant]{This work was presented at the XXXI Marian Smoluchowski Symposium on Statistical Physics in Zakopane (Poland) on September 2nd--7th 2018. The financial support by the grant of The Ministry of Education, Science, Research and Sport of the Slovak Republic under the contract No. VEGA 1/0043/16 and by the grant of Slovak Research and Development Agency under contract No.~APVV-0097-14 is gratefully acknowledged.}

\begin{abstract}
The mixed spin-(1/2, 3/2) Ising model on a decorated square lattice, which takes into account lattice vibrations of the spin-3/2 decorating magnetic ions at a quantum-mechanical level under the assumption of a perfect lattice rigidity of the spin-1/2 nodal magnetic ions, is examined via an exact mapping correspondence with the effective spin-1/2 Ising model on a square lattice. Although the considered magnetic structure is in principle unfrustrated due to bipartite nature of a decorated square lattice, the model under investigation may display anomalous spin frustration driven by a magnetoelastic coupling. It turns out that the magnetoelastic coupling is a primary cause for existence of the frustrated antiferromagnetic phases, which exhibit a peculiar coexistence of antiferromagnetic long-range order of the nodal spins with a partial disorder of the decorating spins with possible reentrant critical behaviour. Under certain conditions, the anomalous spin frustration caused by the magnetoelastic coupling is responsible for unprecedented absence of spontaneous long-range order in the mixed-spin Ising model composed from half-odd-integer spins only. 
\end{abstract}


\end{frontmatter}

\section{Introduction}

Two-dimensional Ising models afford paradigmatic examples of exactly solved lattice-statistical models, which may exhibit nontrivial phase transitions and critical points at finite temperatures \cite{mco73,lav99,str15}. It should be emphasized, however, that the Ising models are mostly defined on perfectly rigid two-dimensional lattices in order to preserve their exact tractability, which means that vibrational degrees of freedom of the magnetic ions underlying a persistent quantum-mechanical motion are frequently completely ignored. Of course, the magnetoelastic coupling between magnetic and lattice degrees of freedom is pervasive in nature and not always negligible with respect to the superexchange coupling, which represents the leading-order interaction term in most real-world insulating magnetic materials \cite{kra34,and50}. 
   
The magnetoelastic coupling between magnetic and lattice degrees of freedom of two-dimensional Ising models can be treated at different levels. On the one hand, one may take advantage of diverse phenomenological theories \cite{bea62,ber76,hen87,mas00,bou01,pli10,bal17,sza18,bal18}, which combine different approximations for individual subsystems such as mean-field or effective-field approximation for a magnetic subsystem, Einstein or Debey approximation for a lattice subsystem, etc. The main advantage of the phenomenological theories lies in their general validity, easy applicability and adaptability, which are however balanced by an incapability to provide insight at the microscopic level. On the other hand, the microscopic lattice-statistical models taking into consideration the magnetoelastic coupling cannot be treated rigorously on account of mathematical complexities closely connected with a coupling between the magnetic and lattice degrees of freedom.  

Recently, we have introduced and rigorously solved a mixed spin-(1/2, $S$) Ising model on decorated planar lattices, which are composed from spin-1/2 nodal magnetic ions  placed at rigid lattice positions and spin-$S$ decorating magnetic ions capable of quantum-mechanical vibrations \cite{str12,str19}. Although the formulation and solution of the mixed spin-(1/2, $S$) Ising model on decorated planar lattices partially liable to vibrations as presented in Ref. \cite{str19} is quite general, the numerical results were up to now presented just for two particular cases having the quantum spin number $S=1/2$ \cite{str12} and $S=1$ \cite{str19}, respectively. In the present Letter, we will therefore thoroughly examine another particular case of the mixed spin-(1/2, 3/2) Ising model on a decorated square lattice, which contains at relaxed decorating lattice sites magnetic ions with the higher half-odd-integer spin $S=3/2$. 
 
This Letter is organized as follows. The model and basic steps of the calculation procedure are reviewed in Section \ref{model}. The most interesting results for the ground-state and finite-temperature phase diagrams are presented in Section \ref{result} together with typical temperature dependencies of the spontaneous uniform and staggered magnetizations. Finally, some conclusions and future outlooks are mentioned in Section \ref{conclusion}. 

\section{Model}
\label{model}

\begin{figure}
\begin{center}
\includegraphics[width=0.3\textwidth]{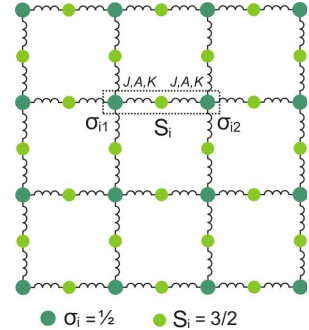}
\end{center}
\vspace{-0.6cm}
\caption{A cross-section from a decorated square lattice. Nodal lattice sites (large dark circles) are occupied by the rigid spin-1/2 magnetic ions (${\sigma}_{i}= \pm 1/2$), while decorating lattice sites (small light circles) are occupied by the spin-$3/2$ magnetic ions ($S_i = \pm 3/2, \pm 1/2$) being subject to lattice vibrations.}
\label{fig1}       
\end{figure}

Let us consider a mixed spin-(1/2, 3/2) Ising model on a decorated square lattice displayed in Fig.~\ref{fig1}, which is constituted by the spin-$1/2$ magnetic ions placed at rigid nodal lattice sites and the spin-3/2 magnetic ions performing instantaneous quantum-mechanical motion at decorating lattice sites. Under these assumptions, the Hamiltonian of the mixed spin-(1/2, 3/2) Ising model on a decorated square lattice can be defined as a sum over bond Hamiltonian ${\hat{\cal H}}_{i}$ 
\begin{eqnarray}
{\hat{\cal H}}\!\!\!&=&\!\!\!\sum_{i=1}^{2N}{\hat{\cal H}}_{i}=\sum_{i=1}^{2N}({\hat{\cal H}}_{i}^{m}+{\hat{\cal H}}_{i}^{e}),
\label{ha}
\end{eqnarray}
which involves all interaction terms of the $i$th decorating magnetic ion. The bond Hamiltonian ${\hat{\cal H}}_{i}$ is further split into the magnetoelastic part 
\begin{eqnarray}
{\hat{\cal H}}_{i}^{m}\!\!\!&=&\!\!\!-\left(J-A\hat{\rho}_{i}\right)S_{i}{\sigma}_{i1}
                                     -\left(J+A\hat{\rho}_{i}\right)S_{i}{\sigma}_{i2}-DS_{i}^2,
\label{ham}
\end{eqnarray}
and the pure elastic part 
\begin{eqnarray}
{\hat{\cal H}}_{i}^{e}\!\!\!&=&\!\!\!\frac{{\hat{p}_{i}^{2}}}{2M}+K\hat{\rho}_{i}^2.
\label{hal}
\end{eqnarray}
The magnetoelastic part (\ref{ham}) of the bond Hamiltonian ${\hat{\cal H}}_{i}^{m}$ takes into account the uniaxial single-ion anisotropy $D$ acting on the decorating spin $S_i$ and the exchange interaction with its two nearest-neighbour spins depending on an instantaneous distance: $J$ is an equilibrium value of the exchange constant, $A$ is magnetoelastic coupling constant and $\hat{\rho}_{i}$ is a local coordinate operator assigned to a displacement of the $i$th decorating magnetic ion from its equilibrium lattice position. The pure elastic part (\ref{hal}) of the bond Hamiltonian ${\hat{\cal H}}_{i}^{e}$ involves kinetic energy of the $i$th decorating magnetic ion with the mass $M$ and the elastic energy, which relates to a square of the displacement operator of the $i$th decorating magnetic ion within the harmonic approximation ($K$ is respective bar elastic or spring stiffness constant).

Let us briefly recall only basic steps of the solution of the mixed spin-(1/2, 3/2) Ising model on a decorated square lattice, which has been comprehensively described our previous work \cite{str19}. The magnetic and elastic degrees of freedom can be decoupled through a local canonical coordinate transformation \cite{str12,str19,ent73} 
\begin{eqnarray}
\hat{\rho}_{i}\!\!\!&=&\!\!\!\hat{\rho}_{i}^{\prime}-\frac{A}{2K}S_{i}\left({\sigma}_{i1}-{\sigma}_{i2}\right),
\label{cct}
\end{eqnarray}
which leaves the elastic part of the bond Hamiltonian (\ref{hal}) invariant and eliminates a displacement operator from the magnetoelastic part of bond Hamiltonian (\ref{ham})   
\begin{eqnarray}
{\hat{\cal H}}_{i}^{m \prime}\!\!\!&=&\!\!\!-J S_{i}\left({\sigma}_{i1}+{\sigma}_{i2}\right)
+J^{\prime}S_{i}^{2}{\sigma}_{i1}{\sigma}_{i2}-D^{\prime}S_{i}^{2}.
\label{hamt}
\end{eqnarray}
The magnetoelastic part of the bond Hamiltonian (\ref{hamt}) consequently depends only on the equilibrium value of the pair exchange constant $J$, the effective three-site four-spin interaction $J^{\prime}=A^{2}/2K$ and the rescaled uniaxial single-ion anisotropy $D^{\prime}=D+A^{2}/8K$. It turns out that the three-site four-spin interaction  $J^{\prime}=A^{2}/2K$ may be responsible for anomalous spin frustration, which arises from a competition with the equilibrium value of the bilinear exchange constant $J$ \cite{str19,jas16,stu17}. 

The bond partition function corresponding to the magnetoelastic part of the bond Hamiltonian (\ref{hamt}) can be substituted through the generalized decoration-iteration transformation \cite{fis59,syo72,roj09,str10,roj11}, which establishes an exact mapping correspondence between the partition function of the mixed spin-(1/2, 3/2) Ising model on a decorated square lattice accounting for lattice vibrations of the decorating magnetic ions and the partition function of the effective spin-1/2 Ising model on a perfectly rigid square lattice with the temperature-dependent nearest-neighbour interaction $\beta R_1 = 2 \ln(V_1/V_2)$
\begin{eqnarray}
{\cal Z}\!\!\!&=&\!\!\!\left[\frac{V_{1}V_{2}}{4\sinh^{2}\left(\frac{\beta\hbar\omega}{2}\right)}\right]^{N} 
{\cal Z}_{\rm IM}\left(\beta R_{1}\right).
\label{mr}
\end{eqnarray} 
For completeness, let us explicitly quote definitions for the expressions $V_1$ and $V_2$
\begin{eqnarray}
V_{1}\!\!\!&=&\!\!\!\sum_{n=-3/2}^{3/2}\exp\left(\beta Dn^2\right)\cosh\left(\beta Jn\right), \nonumber \\
V_{2}\!\!\!&=&\!\!\!\sum_{n=-3/2}^{3/2}\exp\left[\beta n^2\left(D+\frac{A^2}{4K}\right)\right].
\label{v}
\end{eqnarray}
An exact result for the partition function of the spin-1/2 Ising model on a square lattice has been derived in a seminal paper by Onsager 
\cite{ons44} and hence, the mapping relationship (\ref{mr}) provides an exact result for the partition function of the mixed spin-(1/2, 3/2) Ising model on a decorated square lattice being subject to lattice vibrations of the decorating magnetic ions. In addition, one may also derive similar mapping relations for other quantities of particular interest. The critical boundaries of the mixed spin-(1/2, 3/2) Ising model on a decorated square lattices can be for instance calculated from a comparison of the effective coupling $\beta R_1 = 2 \ln(V_1/V_2)$ with its critical value $\beta_{C} |R_1| = 2 \ln (1 + \sqrt{2})$. The readers interested in further calculation details are referred to Ref. \cite{str19}.

\section{Results and discussions}
\label{result}

Let us proceed to a discussion of the most interesting results for the mixed spin-(1/2, 3/2) Ising model on a decorated square lattice, which takes into account lattice vibrations of the decorating spin-3/2 magnetic ions under the assumption of a perfect lattice rigidity of the nodal spin-1/2 magnetic ions. It should be mentioned that the magnetic behaviour of the model under investigation does not basically depend on whether the equilibrium value of the exchange constant is being ferromagnetic ($J>0$) or antiferromagnetic ($J<0$), because the respective sign change merely causes change in a relative orientation of the nodal and decorating magnetic ions but it does not affect absolute values of the critical temperature and sublattice magnetizations. Accordingly, the absolute value of the exchange constant $|J|$ will  henceforth serve as an energy unit. 

\begin{figure}
\begin{center}
\includegraphics[width=0.5\textwidth]{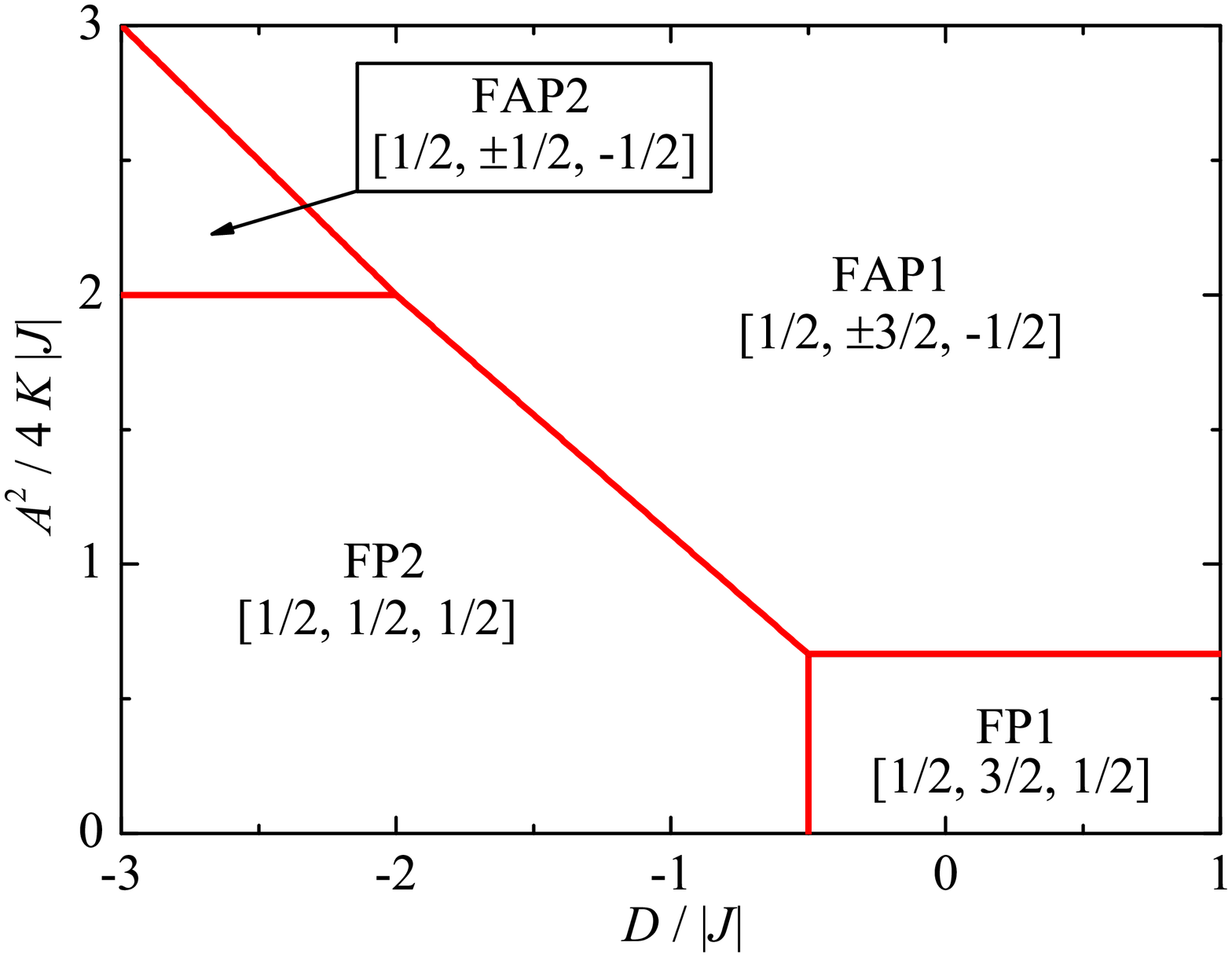}
\end{center}
\vspace{-0.9cm}
\caption{Ground-state phase diagram in the $D/|J|-A^{2}/4K|J|$ plane. The notation for individual ground states is as follows: FP1 (FP2) - ferromagnetic or ferrimagnetic phase 1 (2), FAP1 (FAP2) - frustrated antiferromagnetic phase 1 (2). The displayed phase diagram is invariant under the sign change $J \to -J$, which merely causes trivial change in a relative orientation of the nearest-neighbour spins. The numbers in square brackets determine a typical spin configuration within one bond of each ground state for the particular case with the ferromagnetic coupling constant $J>0$.}
\label{fig2}       
\end{figure}

The ground-state phase diagram, which is depicted in Fig. \ref{fig2} in the $D/|J|-A^{2}/4K|J|$ plane, consists of four different phases. More specifically, one encounters 
the ferromagnetic (ferrimagnetic) phase 1 (FP1)
\begin{eqnarray}
|{\rm FP1} \rangle = \prod_{i=1}^N  \left| {\rm sign}(J) \frac{1}{2} \right \rangle_{\sigma_i}  
                     \prod_{i=1}^N \left| \frac{3}{2} \right \rangle_{S_i}, 
\label{fp1}
\end{eqnarray}
the ferromagnetic (ferrimagnetic) phase 2 (FP2)
\begin{eqnarray}
|{\rm FP2} \rangle = \prod_{i=1}^N  \left| {\rm sign}(J) \frac{1}{2} \right \rangle_{\sigma_i}  
                     \prod_{i=1}^N \left| \frac{1}{2} \right \rangle_{S_i}, 
\label{fp2}
\end{eqnarray}
the frustrated antiferromagnetic phase 1 (FAP1)
\begin{eqnarray}
|{\rm FAP1} \rangle = \prod_{i=1}^{N/2} \left| - \frac{1}{2} \right \rangle_{\sigma_{i1}}  
                      \prod_{i=1}^{N/2} \left| \frac{1}{2} \right \rangle_{\sigma_{i2}}
                      \prod_{i=1}^{N} \left| \pm \frac{3}{2} \right \rangle_{S_i}, 
\label{fap1}
\end{eqnarray} 
and the frustrated antiferromagnetic phase 2 (FAP2)
\begin{eqnarray}
|{\rm FAP2} \rangle = \prod_{i=1}^{N/2} \left| - \frac{1}{2} \right \rangle_{\sigma_{i1}}  
                      \prod_{i=1}^{N/2} \left| \frac{1}{2} \right \rangle_{\sigma_{i2}}
                      \prod_{i=1}^{N} \left| \pm \frac{1}{2} \right \rangle_{S_i}. 
\label{fap2}
\end{eqnarray} 
The ground states FP1 and FP2 exhibit a standard ferromagnetic or ferrimagnetic long-range order depending on a character of the coupling constant $J>0$ or $J<0$, whereas they differ from one another through a magnetic moment of the decorating spin-3/2 magnetic ions being fully saturated within the ground state FP1 and unsaturated within the ground state FP2. The similar difference can be also found in two unconventional ground states FAP1 and FAP2, which display a peculiar disorder of the decorating spin-3/2 magnetic ions on account of a spin frustration originating from a perfect N\'eel (antiferromagnetic) long-range order of the nodal spin-1/2 magnetic ions. The ground states FAP1 and FAP2 thus exhibit a remarkable combination of the perfect long-range order of the nodal spin-1/2 magnetic ions accompanied with a partial disorder of the decorating spin-3/2 magnetic ions.

\begin{figure}
\begin{center}
\includegraphics[width=0.5\textwidth]{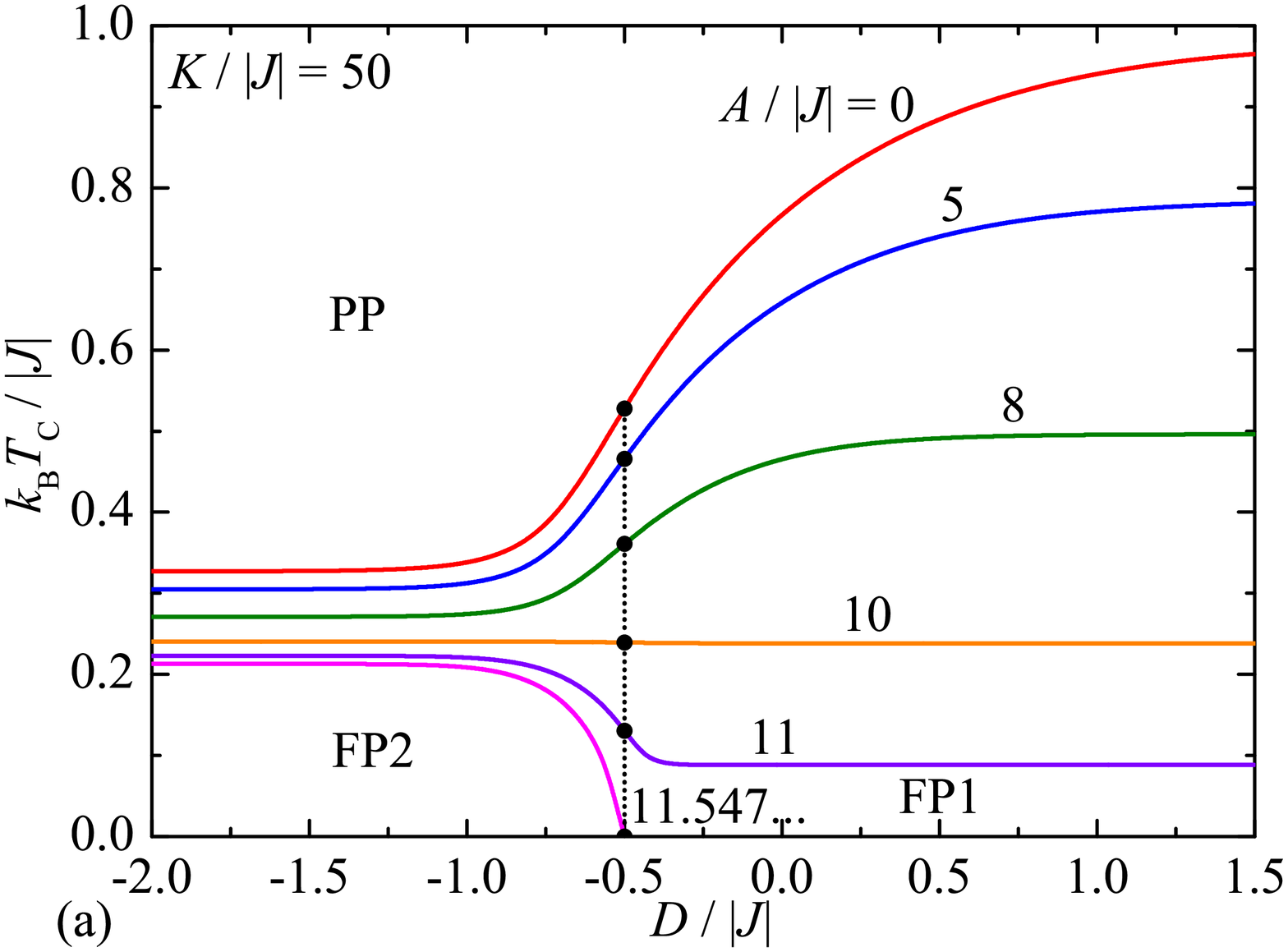}
\vspace*{-0.2cm}
\includegraphics[width=0.5\textwidth]{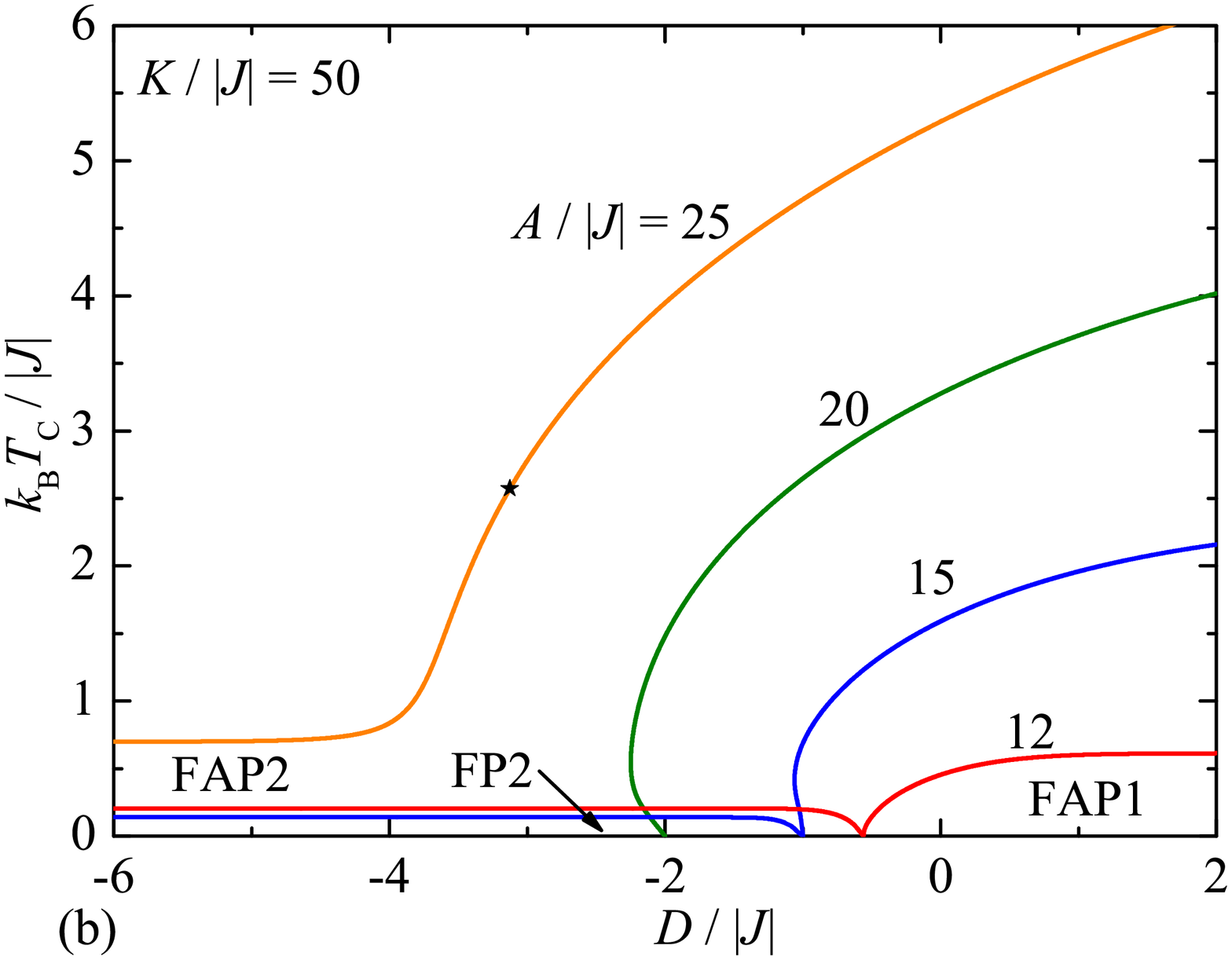}
\end{center}
\vspace{-0.9cm}
\caption{Critical temperature versus uniaxial single-ion anisotropy $D/|J|$ plot for the fixed value of the bare elastic constant $K/|J| = 50$ and several values ​​of the magnetoelastic coupling constant $A/|J|$. Black dots in Fig. \ref{fig3}(a) delimit a part of the critical line pertinent to the FP1 (right) and FP2 (left) ground states. A star in Fig. \ref{fig3}(b) delimits a part of the critical line pertinent to the FAP1 (right) and FAP2 (left) ground states.}
\label{fig3}       
\end{figure}

Now, let us explore a critical behaviour of the mixed spin-(1/2, 3/2) Ising model on a decorated square lattice as a function of the uniaxial single-ion anisotropy $D/|J|$ and the magnetoelastic coupling constant $A/|J|$. First, we will consider sufficiently small values of the magnetoelastic coupling constant $A<\sqrt{8K|J|/3}$, which may invoke according to the ground-state phase diagram shown in Fig. \ref{fig2} either FP1 or FP2 ground state depending on a relative size of the uniaxial single-ion anisotropy $D/|J|$ [see Fig. \ref{fig3}(a)]. It surprisingly turns out that a typical monotonic depression of the critical temperature achieved upon decreasing of the anisotropy parameter $D/|J|$ can be found only for small enough values of the magnetoelastic coupling constant (e.g. $A/|J| < 10$ for $K/|J| = 50$). In fact, the critical temperature becomes almost completely independent of the uniaxial single-ion anisotropy $D/|J|$ at moderate values of the magnetoelastic coupling constant ($A/|J| \approx 10$ for $K/|J| = 50$) or even unexpectedly rises for stronger values of the magnetoelastic coupling constant ($A/|J| > 10$ for $K/|J| = 50$) upon lowering of the anisotropy term $D/|J|$. Most strikingly, the spontaneous long-range order is completely absent for the range of uniaxial single-ion anisotropies $D/|J|>-1/2$ and the particular value of the magnetoelastic coupling constant $A=\sqrt{8K|J|/3}$ (i.e. $A/|J|=20/\sqrt{3} \approx 11.547...$ for $K/|J|=50$). The absence of a spontaneous long-range order in the mixed spin-(1/2, 3/2) Ising model on a decorated square lattice can be attributed to the anomalous spin frustration caused by the magnetoelastic coupling, which is responsible for a phase coexistence of the ground states FP1 and FAP1 with two competing (ferromagnetic vs. antiferromagnetic) orders.

An outstanding criticality of the mixed spin-(1/2, 3/2) Ising model on a decorated square lattice can be also found at stronger values of the magnetoelastic coupling constant $A>\sqrt{8K|J|/3}$, which may favour according to the ground-state phase diagram shown in Fig. \ref{fig2} two unconventional ground states FAP1 and FAP2 apart from the conventional ground state FP2 [see Fig. \ref{fig3}(b)]. In agreement with general expectations, the critical frontiers pertinent to the FAP1 and FP2 ground states meet at the relevant ground-state phase boundary when the magnetoelastic coupling constant is taken from the interval $A \in (\sqrt{8K|J|/3}, \sqrt{8K|J|})$. It should be nevertheless mentioned that the critical line pertinent to the FAP1 ground state may slightly intervene at higher temperatures above the parameter space, which is consistent with the FP2 ground state at zero temperature. Hence, it follows that the mixed spin-(1/2, 3/2) Ising model on a decorated square lattice may display reentrant phase transitions, which relate to emergence of the partially ordered and partially disordered FAP1 phase at higher temperatures. Note furthermore that unprecedented absence of the spontaneous long-range order may be again detected for the range of uniaxial single-ion anisotropies $D/|J|<-2$ and the particular value of the magnetoelastic coupling constant $A=\sqrt{8K|J|}$ (i.e. $A/|J|=20$ for $K/|J|=50$), which in combination cause a phase coexistence of the ground states FP2 and FAP2 with two competing (ferromagnetic vs. antiferromagnetic) orders. Last but not least, one recovers ordinary monotonic fall of the critical temperature upon decreasing of the anisotropy parameter $D/|J|$ for sufficiently high values of the magnetoelastic coupling constant $A>\sqrt{8K|J|}$ (e.g. $A/|J| > 20$ for $K/|J| = 50$), which is this time connected to a phase transition between the FAP1 and FAP2 ground states (the star symbol shown in Fig. \ref{fig3}(b) for $A/|J| = 25$). 

\begin{figure}
\begin{center}
\includegraphics[width=0.5\textwidth]{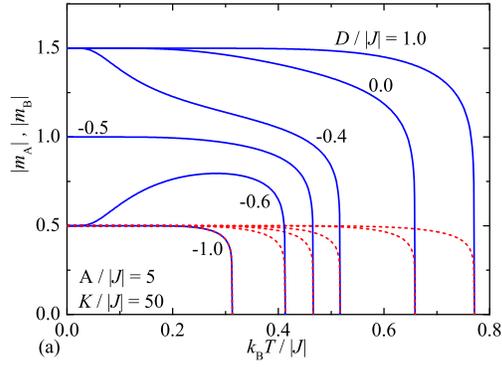}
\vspace*{-0.2cm}
\includegraphics[width=0.5\textwidth]{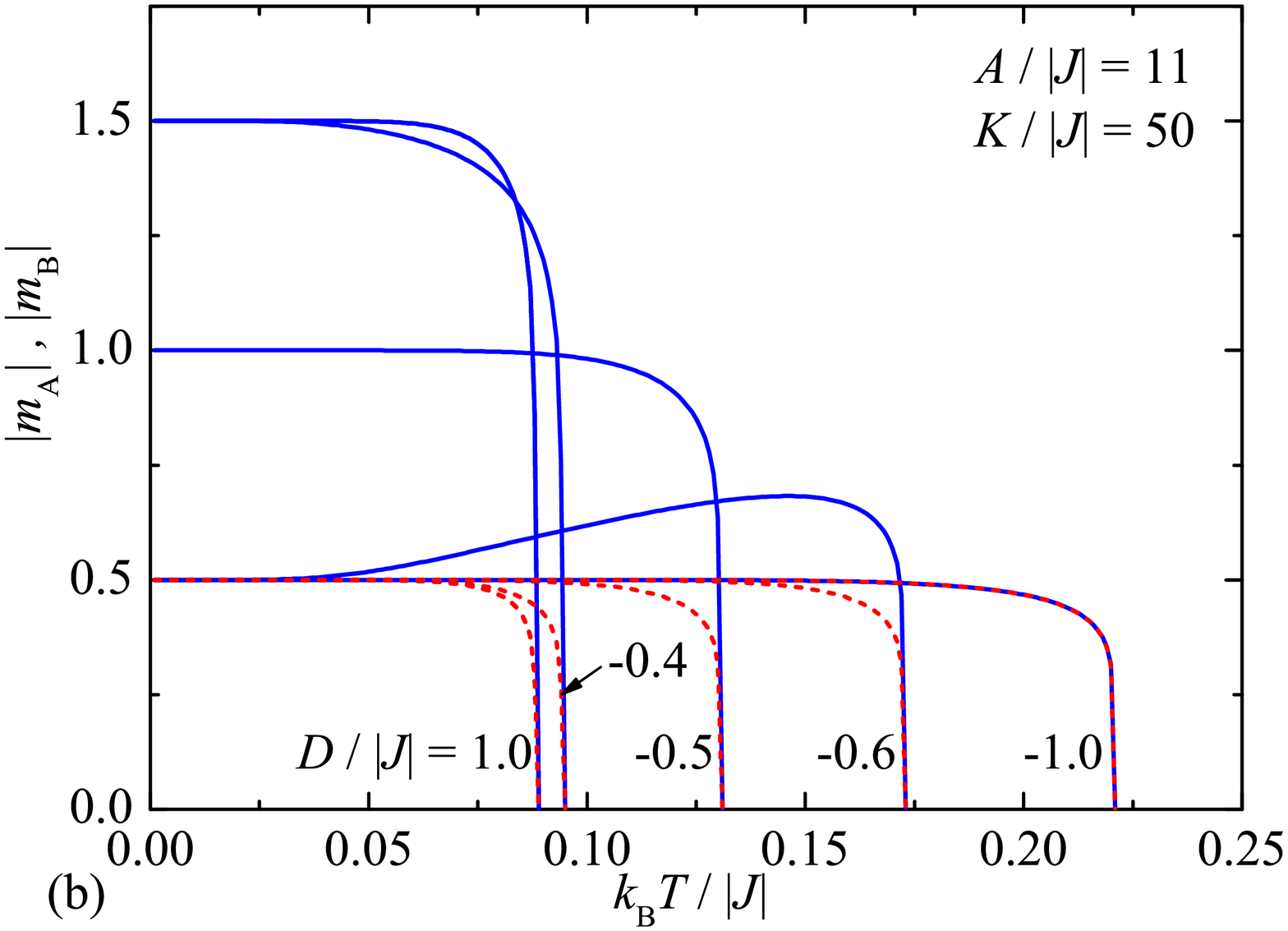}
\end{center}
\vspace{-0.9cm}
\caption{Temperature dependencies of the spontaneous sublattice magnetizations $|m_A| = |\langle (\sigma_{i1} + \sigma_{i2})/2 \rangle|$ and $|m_B| = |\langle S_i \rangle|$ of the nodal and decorating spins for the fixed value of the bare elastic constant $K/|J|=50$, several values ​​of the uniaxial single-ion anisotropy $D/|J|$ and two different values of the magnetoelastic coupling constant: 
(a) $A/|J|=5$; (b) $A/|J|=11$.}
\label{fig4}       
\end{figure}

To verify correctness of the aforedescribed ground-state and finite-temperature phase diagrams of the mixed spin-(1/2, 3/2) Ising model on a decorated square lattice we will comprehensively examine a thermal behaviour of the uniform and staggered sublattice magnetizations serving as the relevant order parameters for a spontaneous ferromagnetic and antiferromagnetic long-range orders, respectively. Temperature dependencies of the absolute values of the spontaneous sublattice magnetizations $|m_A| = |\langle (\sigma_{i1} + \sigma_{i2})/2 \rangle|$ and $|m_B| = |\langle S_i \rangle|$ of the nodal and decorating spins plotted in Fig. \ref{fig4} corroborate a phase transition between two conventional ground states FP1 and FP2 whenever the magnetoelastic coupling constant is sufficiently small $A<\sqrt{8K|J|/3}$. As far as a mutual effect of the magnetoelastic coupling and uniaxial single-ion anisotropy upon a relative size of the critical temperature is concerned, the displayed temperature variations of the spontaneous sublattice magnetizations $|m_A|$ and $|m_B|$ serve in evidence of two markedly different tendencies when a relevant decrease in the anisotropy parameter parameter $D/|J|$ reduces (enhances) the critical temperature at smaller (greater) values of the magnetoelastic coupling constant [c.f. Fig. \ref{fig4}(a) and (b)].

\begin{figure}
\begin{center}
\includegraphics[width=0.5\textwidth]{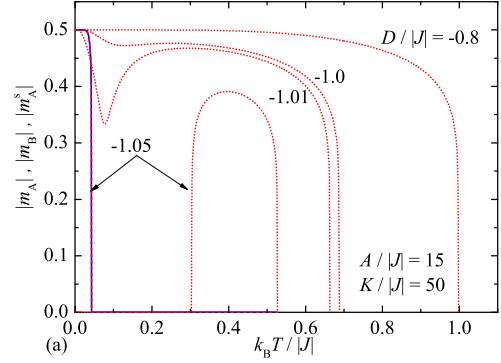}
\vspace*{-0.2cm}
\includegraphics[width=0.5\textwidth]{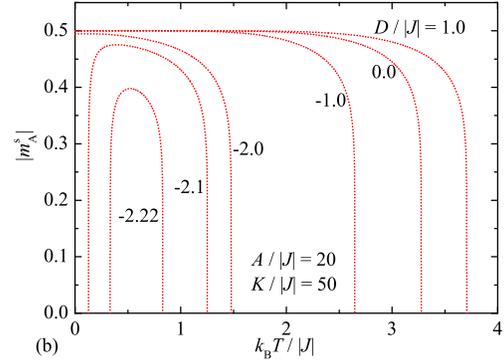}
\end{center}
\vspace{-0.9cm}
\caption{Temperature dependencies of the staggered magnetization $|m_A^s| = |\langle (\sigma_{i1} - \sigma_{i2})/2 \rangle|$ of the nodal spins (dotted lines) for the fixed value of the bare elastic constant $K/|J|=50$, several values ​​of the uniaxial single-ion anisotropy $D/|J|$ and two different values of the magnetoelastic coupling constant: (a) $A/|J|=15$; (b) $A/|J|=20$. The particular case with the uniaxial single-ion anisotropy $D/|J|=-1.05$ shown in Fig. \ref{fig5}(a) displays at low enough temperatures nonzero sublattice magnetizations $|m_A| = |\langle(\sigma_{i1} + \sigma_{i2})/2 \rangle|$ and $|m_B| = |\langle S_i \rangle|$ of the nodal and decorating spins (solid lines).}
\label{fig5}       
\end{figure}

Finally, let us turn our attention to the mixed spin-(1/2, 3/2) Ising model on a decorated square lattice with a strong enough magnetoelastic coupling $A>\sqrt{8K|J|/3}$, which may display a peculiar coexistence of a spontaneous long-range order of the nodal spins with a partial disorder of the decorating spins within the ground states FAP1 and FAP2. To this end, we have plotted in Fig. \ref{fig5} typical temperature dependencies of the spontaneous staggered magnetization $|m_A^s| = |\langle (\sigma_{i1} - \sigma_{i2})/2 \rangle|$ of the nodal spins for the set of parameters related to the ground state FAP1. It is quite evident from Fig. \ref{fig5}(a) that the more negative the uniaxial single-ion anisotropy $D/|J|$ is, the smaller is the critical temperature associated with disappearance of the spontaneous staggered magnetization. It should be emphasized, moreover, that the staggered magnetization of the nodal spins exhibits notable temperature variations close to a phase boundary between the ground states FAP1 and FP2. As a matter of fact, the staggered magnetization of the nodal spins displays either a nonmonotonous temperature dependence with a pronounced deep minimum at moderate temperature [see $D/|J| = -1.01$ in Fig. \ref{fig5}(a)] or with a double reentrant critical behaviour [see $D/|J| = -1.05$ in Fig. \ref{fig5}(a)]. Note that the reentrant dome of the staggered magnetization $|m_A^s|$ with two consecutive critical points already appears for $D/|J| = -1.05$ above the ground state FP2 with a spontaneous ferromagnetic long-range order as evidenced by nonzero spontaneous uniform magnetizations $|m_A|$ and $|m_B|$ of the nodal and decorating spins. It is worthwhile to remark that the temperature dependence with a double reentrant behavior of the spontaneous staggered magnetization $|m_A^s|$ of the nodal spins may come into being also without a prior existence of the low-temperature ferromagnetic ground state FP2 as exemplified in Fig. \ref{fig5}(b). 

\section{Concluding remarks}
\label{conclusion}

In the present work we have investigated in detail magnetic and critical properties of the mixed spin-(1/2, 3/2) Ising model on a decorated square lattice, which takes into account lattice vibrations of the decorating spin-3/2 magnetic ions at a quantum-mechanical level under the simultaneous assumption of a perfect lattice rigidity of the nodal spin-1/2 magnetic ions. It has been demonstrated that the magnetoelastic coupling gives rise to an effective three-site four-spin interaction competing with an equilibrium value of the pair spin-spin interaction, which leads to anomalous spin frustration arising in spite of a bipartite (unfrustrated) nature of a decorated square lattice. In particular, our attention has been paid to a detailed study of the ground-state and finite-temperature phase diagrams together with temperature dependencies of the spontaneous uniform and staggered magnetizations.

It has been convincingly evidenced that the magnetoelastic coupling is a primary cause of two frustrated antiferromagnetic ground states FAP1 and FAP2 with an outstanding coexistence of a perfect N\'eel (antiferromagnetic) long-range order of the nodal spin-1/2 magnetic ions with a partial disorder of the decorating spin-3/2 magnetic ions of either fully saturated (FAP1) of unsaturated nature (FAP2). It turns out, moreover, that the remarkable coexistence of the antiferromagnetic order of the nodal spins with a partial disorder of the decorating spins is an essential ingredient for observing a double reentrant critical behaviour. Most strikingly, the anomalous spin frustration may also cause a complete absence of spontaneous long-range order whenever the magnetoelastic coupling drives the mixed spin-(1/2, 3/2) Ising model on a decorated square lattice to a phase boundary between two ground states with the competing (ferromagnetic vs. antiferromagnetic) spin orders. To the best of our knowledge, the lack of criticality in a two-dimensional mixed-spin Ising model composed solely from half-odd-integer spins has been not reported hitherto.

\end{document}